# 1-d granular gas with little dissipation in 0-g :

# A comment on "Resonance oscillations in Granular gases"

## P. Evesque

Lab MSSMat, UMR 8579 CNRS, Ecole Centrale Paris
92295 CHATENAY-MALABRY, France, e-mail: evesque@mssmat.ecp.fr

### Abstract:

*It is demonstrated that recent results on 1d granular gas in a container with a vibrating piston, which was modelled by a shock wave propagation, can be understood with a modelling using ideas coming from the "thermodynamics of a single particle". Defining e as the square root of the energetic-restitution coefficient of a single collision, the mean loss during a round trip of the momentum is calculated in the limit* $N(1-e)\ll 1$. *It is also demonstrated that the system cannot propagate sound waves nor shock waves in the limit of* $N(1-e)\ll 1$ *and that hydrodynamics equations cannot be defined when* $N(1-e)\ll 1$.

**Pacs # :** 5.40 ; 45.70 ; 62.20 ; 83.70.Fn

___________________________________________________________________

Recently, hydrodynamics models using shock waves propagation have been proposed [1] to model and understand the dynamics of a 1d granular gas of N identical particles contained in a 1d closed sample with one wall vibrating sinusoidaly as a function of time t (frequency f , amplitude A), *i.e.* $x = A \cos(2\pi f t)$. In the following paper, we try and understand the same problem when the losses are small; in this case it is demonstrated that the approximation of "the thermodynamics of a single bead in a 1d container" proposed recently [2] shall be valid under the hypotheses of [1] , *i.e.* (i) the balls are identical, (ii) the restitution coefficient with the wall is 1 and between beads is $e^2$.

Results in [1] are expressed in terms of the typical kinetic energy $E_T=3.6(Lf)^2$ and of the dissipative parameter $D=N(1-e)$. On the contrary, the present paper shows that the "single-particle modelling" works quite well as long as the dissipation due to collisions remains small, *i.e.* $N(1-e)\ll 1$, where e relates the relative speeds before ($\Delta v$) and after ($\Delta v'$) the collision [1] , *i.e.* $\Delta v'=-e\Delta v$. It is found also that the right parameter governing the kinetic energy is $v^2\approx(Af)^2/D$ instead of $(Lf)^2$ as it was proposed in [1].

Furthermore the paper demonstrates that neither hydrodynamics equation nor pressure wave can be defined in a 1d mono-atomic gas so that shock waves cannot exist, strictly speaking. This is then incompatible with the understanding developed in [1].

The paper starts and investigates the case of perfectly elastic collisions (e=1). It shows in this case that the problem is strictly equivalent to the problem of the thermodynamics of a single bead in a 1d container. It continues with the case e slightly different from 1 for which the losses are calculated. Then problems of sound wave propagation and of hydrodynamics equations are considered and some results on Knudsen regime are recalled. This allows demonstrating that one cannot define the average quantities requested in order to define the physics of wave propagation and the hydrodynamics limit.

## 1. The case e=1

Let us start first by considering the problem with a restitution coefficient e=1. In this case the collision between two particles i and j results in the interchange of the two speeds: calling $(v_i, v_j)$ and $(v'_i, v'_j)$ the speeds of the particles before and after the collision, the collision equations, *i.e.* momentum conservation equation and total energy conservation (since e=1), write $(v_i+v_j) = (v'_i+v'_j)$ and $v_i^2+v_j^2 = v'^2_i+v'^2_j$ and impose $v_i = v'_j$ and $v_j = v'_i$. Interchanging the labels i↔j after each collision (i,j) allows to write that each particle number i continues its trajectory with the same speed as if it was not collided with the other one. So, the problem of N identical rigid non dissipating beads in an excited 1d container is then equivalent to N uncorrelated problems of just one bead in an excited 1d container, if the restitution coefficient of the beads is equal to 1. This limit case is then exactly described in paper [2] where it is shown the existence of resonance effect for specific amplitudes and for specific values of the restitution coefficient r, r describing the collision between the ball and the walls. It is worth noting that the right parameters are A/L and r, so that f does not play any role contrarily to what let expect Eqs. (1-3) of paper [1] . Case r=1 has not been studied in [2], since it leads to diverging speed.

However, one can extend the results from r≠1, but r→1 ; in particular, one finds that the speed distribution is peaked around two values $<v_+>$ and $<v_->$ which are related together via r , *cf.* [2] . For the values of r investigated in [2], it has been found also that the relative width $\Delta v/<v_\pm>$ of the distribution is rather constant, *i.e.* $\Delta v/<v_\pm>=\alpha$, with $\alpha \approx 0.3$ [2]. This means that the problem of N identical beads in 1d with no loss during bead-bead collisions and with some losses for collisions with boundaries is then strictly equivalent to the case studied in [2].

This allows proposing a method to visualise the ball trajectories and to understand the complexity of the pattern. This method is reported in Fig. 1. It considers the case of no loss during ball-ball collision (e=1) or during the collision with the top wall ($r_{top}=1$) but some small loss with the moving piston ($r_{bottom}<1$); The method starts with the construction of N independent trajectories of N effective particles bouncing independently in between the two boundaries; such trajectories are starting at different times and are constituted of segment of straight lines which are starting from the bottom and which are reflected by the top fix wall. Then the speed of these effective particles change randomly when they meet the moving bottom wall; in the limit of v<<Lf and of v>>2πAf, the variation δv of this speed is rather random in the range -2πAf <δv<2πAf , *cf.* [2]. So, the trajectories of these effective particles are made of a series of connected Λ whose branches are symmetric and more or less inclined.

As exemplified in Fig. 1, one can find the real trajectory of the balls by continuity and using the following rules: (i) the balls follow the effective-particle trajectories; (ii) the intersection of two of these effective-particle trajectories represents a collision between adjacent balls. An example of real-ball trajectory is given in Fig. 1.c.

Also the method can be adapted to solve the problem of a column of N balls bouncing under gravity, as mentioned in the Fig. 1 caption.

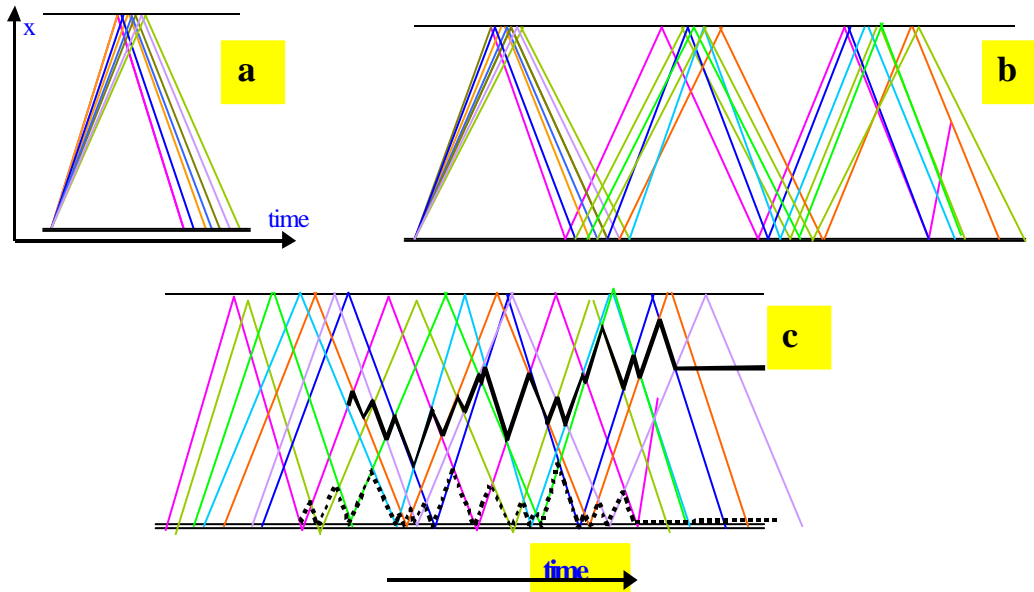

*Figure 1: Dynamical evolution of a 1d column of bouncing balls between a non moving wall (top horizontal black line) and a vibrating piston (bottom black double line); the axis x of the ball position is vertical; the amplitude of vibration A is supposed to be small compared to the cell size L, i.e. A/L<<1 and $v^+$<<Lf , so that the RPA approximation is satisfied [2].*

*Fig. 1.a represents different typical round trips for a single bead with r=1 for the non-moving wall and r≈1 for the piston. Slope of the trajectory defines the speed; under these conditions $v^+$=$v^-$ after the bouncing on the top wall.*

*Fig. 1.b: A real trajectory of a bouncing ball is any combination of connected round trips, such as $v^+_n$-2πAf <$v^+_{n+1}$<$v^+_n$+2πAf. 7 different possible trajectories are represented with 7 different colors, each of them starting from the same origin of time.*

*Fig. 1.c represents a typical time evolution of a column of N bouncing balls with e=1, $r_{top}$=1, $r_{bottom}$<1, (with N=7 here). It is constituted by N independent trajectories of N single effective bodies, each trajectory has a different colour; they starts all from different times. The intersections of two different trajectories represent the collisions between adjacent bouncing balls. Real trajectories of real bouncing balls can be found by following the series of impacts. Two real trajectories of real balls have been represented, the first one is the bottom ball and is in thick dashed black line, the other is in the middle of the column and is in thick continuous black lines.*

*This method can be extrapolated to the problem of N bouncing balls under gravity; in this case each effective trajectories which are here segments of straight line which start from the bottom (top) boundary and end at the top (bottom) boundary would be replaced by an arch of parabola.*

## 2. The case e≠1, but N(1-e)<<1

When e≠1, each collision generates some energy loss; however, the losses are small till $e^{2N}$≈1. So, one can solve the problem starting from the e=1 solution using a perturbation theory when N(1-e)≈1. In particular, one can follow the evolution of a pseudo-particle bearing at time $t_1$ a momentum (or speed) $v_1$. In the last section, it has been proposed that the continuity of the pseudo trajectory is found by interchanging the ball numbers i↔j after each (i,j) collision; we follow the same procedure here. So, the pseudo particle coincides with ball 1 at $t_1$, but its speed v evolves after n collisions and its trajectory coincide now with the one of ball n+1. This leads to conclude that the pseudo particle (and its speed) propagates from ball to ball till it comes back after enough (*i.e.* ≈2N-1) collisions; so, the speed $v_1$ at time t of the ball number 1 located near the moving wall comes back with the value $-v'_1$, *i.e.* in the other direction, after a delay $\delta t \approx 4L/(v_1-v'_1)$ and after 2(N-1) collisions with the other balls plus a collision with the non moving wall. Each collision between balls (i,j) satisfy the equations:

$$v'_i + v'_j = v_i + v_j \quad \& \quad v'_i - v'_j = -e(v_i - v_j) \tag{1}$$

which leads to

$$v'_i = (1+e)v_j/2 + (1-e)v_i/2 \quad \& \quad v'_j = (1+e)v_i/2 + (1-e)v_j/2 \tag{2}$$

Noting $\beta = (1+e)/2 \cong 1$ and $\varepsilon = (1-e)/2 \cong 0$, or $e = 1-2\varepsilon$ and $\beta = 1-\varepsilon$, this leads to

$$v'_i = \beta v_j + \varepsilon v_i = (1-\varepsilon)v_j + \varepsilon v_i \quad \& \quad v'_j = \beta v_i + \varepsilon v_j \tag{3}$$

Let us now note $v_p$ the speed of ball p before the collision with ball p-1; this collision ensures the propagation of the speed $v_1$, which was on ball 1 at time $t_1=0$, p-2 collisions before; so one gets after p-1 collisions and after summation of the different equations of the kind of Eq. (3):

$$v'_{n+1} = \beta^n v_1 + \varepsilon[\sum_{i=1}^{n} \beta^p v_p] = (1-\varepsilon)^n v_1 + \varepsilon[\sum_{i=1}^{n} (1-\varepsilon)^p v_{n+1-p}] \tag{4}$$

$v'_{n+1}$ is the speed of ball n+1 at time t between $t_n$ and $t_{n+1}$ which corresponds to the laps of time for which $v_{n+1}$ corresponds to the speed $v_1$ emitted by ball 1 at time $t_1$. By taking the limit $2N\varepsilon \ll 1$, and considering the case of a column of N+1 balls, one gets at second order in $\varepsilon$:

$$v'_{n+1} \cong [1-2N\varepsilon+4N^2\varepsilon^2]v_1 + \varepsilon[\sum_{i=1}^{N} [(1-p\varepsilon)v_p + (1-\{2N-p\}\varepsilon)v'_{p+1}] \tag{5}$$

In Eq. (5), $v'_{p+1}$ is a speed different from $v_{p+1}$, since both speeds are related to two distinct collisions between balls p and p+1 at two different times during a round-trip; so, $v'_p$ means the speed of ball p after a round-trip.

In Eq. (5), the speed $v'_p$ has to be considered as positive when the ball comes back towards the piston. So the convention about the positive direction is changed as soon as particle collides with the top wall: $v_p$ and $v'_p$ are both positive, the first one when the ball moves towards the top wall, the second one when it moves in the opposite direction.

Limiting the calculation to first order and considering a complete round-trip, one gets:

$$v'_1 \cong v_1 + \varepsilon[\sum_{i=1}^{N} [(v_p - v_1) + (v'_{p+1} - v_1)] \tag{6}$$

The value of v'$_1$ depends on 2N random collisions; some of these collisions happens with balls moving in the same way (case C$_A$), others move in opposite ways, (case C$_B$). The second part of the right hand side of Eq. (6) describes the energy (or velocity) losses. Furthermore, the probability of collisions during dt is proportional (v$_p$-v$_1$); it is then much larger when balls are moving in opposite direction. Denoting $\underline{v}$ the mean speed of the other balls, it means that the probability of collisions with balls moving in the same way is rather small if v$_1$>$\underline{v}$; it becomes important only when v$_1$ is smaller than $\underline{v}$. By labelling $\Delta v$ the square root of the standard deviation of $|v_1|-|v_p|$, and by taking the average of Eq. (6), one gets the mean v'$_1$:

$$<v'_1> \cong v_1 - 2N\varepsilon \{[<|v_p|>+|v_1|]^2+(\Delta v)^2\}/\{<|v_p|>+|v_1|+\Delta v\} \qquad (7)$$

In average $\Delta v$ is about the width of the distribution of $|v_p|$, *i.e.* $\Delta v \approx \alpha \underline{v}$ ; in the following we will write $\Delta v \approx \alpha' \underline{v}$. From [2] one knows that $\alpha$=0.3; considering then 2>>$\alpha$:

$$<v'_1> \cong v_1 - 2N\varepsilon \{\underline{v}+v_1-\alpha'\underline{v}\} \qquad (8)$$

Eq. (8) shows that the larger $\alpha'$ the smaller the losses, when $\alpha$ remains small compared to 1; this is simply due to the fact that the number of collisions which occurs between balls going in the same way increases, which results in a decrease of the losses.

Another point is worth noting: the mean loss, *i.e.* v$_1$-<v'$_1$> , as evaluated from Eq. (8), contains a term proportional to v$_1$ and another term proportional to $\underline{v}$. It results from this that the global dynamics of collision is not strictly equivalent to a single-body collision, which is governed by a restitution coefficient which would lead to a relationship of the kind v$_1$-<v'$_1$>=(1-r)v$_1$ . However averaging Eq. (8) over all possible v$_1$ leads to an average rule similar to the one of a single-body collision, since

$$<v'_1> - <v_1> \cong -2(2-\alpha')N\varepsilon \underline{v} \qquad (9)$$

with v$_1$=$\underline{v}$ and $\alpha'$<2$\alpha$.

Another point which is also worth noting is that the fluctuations of v'$_1$ shall be smaller than the one which would results from a single collision with a single particle; this is linked to the rules of statistics which sharpen the relative width of the distribution of an event which is composed by a series of independent events. It imposes then that the distribution of v'$_1$ shall be sharper than the one of v$_1$; however the sharpening shall not be large since $2\alpha N\varepsilon$ is supposed to be small.

♣ real trajectories: One can use the scheme proposed in Fig. 1 to describe the trajectories of the pseudo particles; the only difference comes from the losses which occur at each collision: this reduces the speed of the pseudo-particle and makes the trajectories slightly concave instead of remaining straight; however, this is negligible since 2N(1-e) remains about 1.

It is also worth noting that the pseudo particle model, which is developed here, is a resonant model because each pseudo particle is moving back and forth freely in the box and adjust its speed to the conditions of reflection on the two walls. On the contrary, this does not hold true when the dissipation is too large since each pseudo particle has lost most of its energy in the interactions with the others.

The limit of validity of the pseudo particle model can be estimated using a mean field argument: the loss of speed $v^+_1 - v^-_1$ due to the round trip shall be smaller than the gain the ball can get by collision with the moving piston. However, it does not seem that this argument leads to some incompatibility because such a condition is also imposed in the case of a single particle in a vibrating container [2] for which simulations show that the dynamics of the mean speed of the bead is adjusted spontaneously and obeys this condition whatever the losses [2], even in the limit of $(1-r) \to 0$.

♣ Simplified modelling: Anyhow, when the dissipation due to the 2N-2 collisions of the system of N particles contained in the box is small enough, *i.e.* $N\varepsilon = N(1-e) \ll 1$, it is possible to consider that the system is equivalent to a set of N pseudo particles moving independently and having no interaction with one another; however this supposes that losses are described via the interaction with an effective medium or with a boundary, and that this effective loss obeys the rule given by Eq. (9). In Eq. (9), the $\alpha'$ parameter varies with $|v_1|$ and with $|\underline{v}|$ and $\Delta v$, where $|\underline{v}|$ is the mean speed of the particles and $\Delta v$ the width of the speed distribution; indeed, $\alpha'$ varies from 0 to $2\alpha$ about. So the losses are slightly non linear in $\underline{v} v_1$. However noting that $\Delta v / \underline{v}$ is much smaller than 2 allows to neglect the term $\alpha'$. In this case, Eq. (9) shows that the particle dissipate through a viscous damping $-\eta v$, whose strength $\eta$ is proportional to the number N of beads and is given by $\eta = 4N\varepsilon = 2N(1-e)$.

♣ Mean velocity in the Random Phase Approximation (RPA): one can find the value of the mean speed $\underline{v}$ using a self consistent approximation similar to the one proposed in [2] ; it is based on a random phase approach (RPA) [2]. This states that collision with the moving piston occurs randomly at any phase of the motion with a probability which is proportional to the relative speed of the ball $v'_1 - A\omega \sin(\omega t_{coll})$ in the piston frame, (with $\omega = 2\pi f$); this imposes:

$$<v> = \int_0^T [<v'> + 2 A\omega \sin(\omega t_{coll})] [1 + (A\omega/<v'>)\sin(\omega t_{coll})] dt_{coll}/T$$

$$<v> = \{<v'>^2 + A^2\omega^2\}/<v'> = <v'> + A^2\omega^2/<v'>$$

Combining this last equation with Eq. (9) leads to the self-consistent equation : $2(2-\alpha)N\varepsilon \underline{v} = A^2\omega^2/<v'> \cong A^2\omega^2/\underline{v}$. This leads to

$$\underline{v} = \pi A f/[(1-\alpha/2)N\varepsilon]^{1/2} = 2\pi A f/[(2-\alpha)N(1-e)]^{1/2} \quad \gg 2\pi A f \qquad (10)$$

♣ Resonance condition and mean velocity at resonance: Eq. (10) holds when no resonance occurs and when A/L is small. Resonance effect can occur for specific values of A/L. In this case collisions occur rather periodically when the wall motion has always the same phase defined by $t_{coll}$. In order to describe this resonance, the typical time $t_{coll}$ has to be introduced in the self consistent equation and the integration over a random $t_{coll}$ has to be removed. This leads to the self consistent relation:

$$<v> = <v'> + 2\ A\omega\ \sin(\omega t_{coll}) \qquad (11)$$

Eq. (11) leads to $2(2-\alpha)N\varepsilon\ \underline{v} = 2\ A\omega\ \sin(\omega t_{coll})$. Maximum amplitude is obtain for $\sin(\omega t_{coll})=1$. So the typical speed at resonance is:

$$\underline{v} = 2\pi A f / [(2-\alpha)N\varepsilon] \qquad (12)$$

Neglecting $\alpha$ compared to 2 and noting that resonance condition implies also $2L=\underline{v}/f$ lead to:

$$A/L \approx 2N\varepsilon/\pi \qquad \text{with } \varepsilon=(1-e)/2 \qquad (13)$$

These conditions of resonance, *i.e.* Eqs. (11-13) compare rather well with what has been found in [1].

We discuss now in the next sections the consequences of the above modelling and its consistency with some hydrodynamics modelling and with sound-wave-propagation modelling.

## 3. Sound propagation and 1-d granular gas:

One can ask whether sound waves can propagate in such a medium? As a matter of fact, it is difficult to speak in terms of sound propagation, since this requires to speak in terms of the propagation of average macroscopic quantities like the density or the pressure. Indeed, in the previous sections it has been shown that the excitation of ball 1 at time t propagates through the medium at a velocity equal to the initial speed of ball 1. This has few consequences:

Let us first consider the case of a very short disturbance. In this case the excitation modifies the speed of a unique ball. The disturbance propagates at the speed $v_1$. If one repeats the same disturbance at another moment, the new disturbance propagates with another speed $v'_1$. So $v_1$ and $v'_1$ play the role of the information velocity c, but this velocity varies then from shot to shot and has a typical amplitude of fluctuations which is $\Delta c/c=\alpha \cong 0.3$. Strictly speaking then, it is not possible to speak in terms of wave propagation: two impulses propagate at different speeds so that one which was emitted later can pass the other one which was emitted earlier so that the hierarchy of times is not respected after a while...

Let us now consider the case of a long disturbance. If the disturbance is long, it modifies the initial speed distribution of $v_1$ during a time much longer than $L/(N\underline{v})$. The set of speeds propagate independently, each of them at the speed $c_1 = v_1$ which is a random variable distributed around $\underline{v}$; so $c_1$ is a random variable with a mean fluctuation $\Delta c_1/c_1 \approx \alpha \approx 0.3$ and the signal becomes incoherent quite fast; furthermore, its width broadens linearly with time. Hence there is no coherent propagation.

So the time scale that one has to consider to separate long and short duration of excitations in this experiment is the time $\tau=L/(N\underline{v})$ which is a microscopic time. "Waves" propagate in a coherent manner only if the excitation duration $T_e$ is $T_e <\tau$. But the temporal resolution is not enough to discriminate between composed signal and single signal: both affects the speed of a single particle during a time shorter than the average time between two collisions.

These considerations can be extended to the case of any classical 1d gas of identical mono-atomic particles: let us consider such a gas; due to the conservation of momentum and of energy the dynamics of this 1d gas can be considered as the superposition of the dynamics of N non interacting particles; this needs just to interchange after each collision the numbering of the pair of the two particles having collided. So the particles of such a gas are moving independently from one another and with an infinite mean free path, (or with a mean free path which depends only on the boundary conditions); so, the dynamics of each particle depends only on the dynamics it gets due to the interaction with the boundaries. This gas is then similar to a Knudsen gas [3], which forbids to get averaged local quantities with some physical sense [3], such as pressure, density,…[3] It forbids then to write any equation of sound propagation [3]…

The 1d problem would have been different if the particles were endowed by a series of internal degrees of freedom. Indeed, in this case only global conservation of energy and of momentum shall be written after each collision, which generates some uncertainty and some randomness. If the collisions obey the thermodynamics laws, then the internal degrees of freedom play the role of a thermostat. This is quite sufficient (i) to ensure the validity of the statistics approach, (ii) to ensure the consistency of macroscopic variables and (iii) to ensure the thermodynamics approximation; this ensures then that pressure excitation propagates freely in this medium as sound waves.

Such internal degrees of freedom can exist if the particles are complex and made up of few spheres linked together by springs or if the spheres were elastic but soft. This was not the case considered in the present work for which the spheres were considered to be rigid.

The problem is also different if the space is 2d or 3d, because the momentum is a 2d vector and the result of each collision depends on an internal parameter, *i.e.* the relative angle between the relative direction of motion and the direction of the surface of contact when the two spheres are in contact.

## 4. Hydrodynamics modelling:

In the same way, for the same reasons as those ones given in Section 3, one can not write local dynamics equation similar to those ones of hydrodynamics because these ones are based on local average variables which have no meaning in the present case of a 1d gas of identical particles [3].

## 5. Conclusion:

The dynamics of a 1d granular gas made of mono-disperse rigid spheres has been studied in the case of small losses $N(1-e)<<1$, e being the restitution coefficient at each collision, *i.e.* $\Delta v'=-e\Delta v$. Firstly, it has been demonstrated that the problem is strictly equivalent to N independent problems of a single pseudo-particle in a 1d box when the restitution coefficient e e=1. So this N-body problem coincides with the one of the thermodynamics of a single bead in a 1d container under specific boundary conditions [2]. The phase space dimension of this problem is 2N+1, i.e. 2 for each pseudo particles plus 1 for time, or for the piston motion.

We have shown in Fig. 1 how the study of this 1-body problem can help studying the dynamics of a peculiar ball in a column of N balls bouncing in a vibrated cell.

Secondly, the case of small losses can be approximated within the same scheme of a single particle in a 1d container. Losses have been calculated and the problem solved at first order; it results from this calculation that the loss law for a single particle is rather similar to the one obtained from a viscous damping -ηv, with η being proportional to N(1-e). This regime holds as far as N(1-e)<<1. As mentioned already, higher order terms have not been taken into account so far. They introduce long-range interactions between the effective particles, since N(1-e) <<1.

The case of larger losses, *i.e.* N(1-e)≥1, has not been investigated. The method proposed in this paper cannot be applied in this case because it requires also that the statistics of the speeds $v^{\pm}_p$ of particle p in the two directions +x and -x are always of the same magnitude. This is so in the case studied in this paper because the losses are small after a round trip; but this would not be anymore valid if the losses were large. In that case it would be perhaps possible to try and find a solution for the speed distribution which decreases as an exponential with the particle number n?

As the two first regimes, *i.e.* e=1 or N(1-e)<<1, are equivalent to a single-particle problem, sound wave cannot propagate and hydrodynamics formalism cannot be developed. These two results have been obtained due to the similarity of this N-body problem with the case of very low density gases which is called the Knudsen regime [3]. It turns then out that shock waves cannot be defined in such media and the interpretation proposed in [1] of this problem is probably very weak.

At last it is worth noting that the present modelling, based on a single-particle dynamics, holds true only due to the 1d nature of the dynamics and due to the rigid assumption. The same granular-gas problem in 2d or in 3d, or in 1d with a complex structure of the particles with springs and internal degrees of freedom would have not led to the same conclusion: in these cases, collisions would have generated random and complex changes of momentum. These problems cannot then be reduced to the problem of the dynamics of a single particle with no interaction but are much more complex. It results from this that such systems loose very quickly the memory of their initial conditions so that a mean field theory can be built after few collisions, which allows to reduce the number of independent parameters to the average quantities. These last ones become the main quantities of interests in this case. The larger the space dimension d the better the mean field and the faster the averaging, *i.e.* d=2 or 3.

In this paper it has been shown that the problem of sound propagation is not relevant in a 1d granular gas; this would be different if the particles were connected through springs. It is worth recalling some results about this case: in this case phonon modes can be defined if all masses and all spring strengths are equal. However the theory of Anderson localisation [4] does apply as soon as some disorder exists: for instance it can be demonstrated that all normal modes of vibration extend only over a finite zone which is smaller than the system size in presence of disorder, and the larger the disorder the smaller the extension. This may help understanding that 1d problems are special issues of statistical physics with quite strange behaviours controlled by topological effects.

Finally, the case of a 1d granular gas in a container with mono-disperse particles and with variable restitution coefficient could have been investigated in the same limit, *i.e.* $\Sigma_i(1-e_{i,i+1}) \ll 1$. It would have led to the same conclusions as those developed in the course of this paper with a constant restitution coefficient. Indeed, due to statistical thinning, variable e can be replaced by an effective $\underline{e}$ which is given by $N(1-\underline{e}) = \Sigma_i(1-e_{i,i+1})$.

On the contrary, the case of a 1d granular gas in a container with poly-disperse particles has not been investigated. *A priori*, it is much more intricate because local transfer of momentum depends importantly upon the local mass ratio $m_i/m_{i+1}$ and because the space distribution of this ratio is fix, *i.e.* time invariant. This shall result in long-range-correlation-effect that makes the problem more intricate.

*Acknowledgements:* CNES is thanked for partial funding.

The electronic arXiv.org version of this paper has been settled during a stay at the Kavli Institute of Theoretical Physics of the University of California at Santa Barbara (KITP-UCSB), in june 2005, supported in part by the National Science Fundation under Grant n° PHY99-07949.

*Poudres & Grains* can be found at :
http://www.mssmat.ecp.fr/rubrique.php3?id_rubrique=402